\documentclass[a4paper,twocolumn,11pt,unpublished]{quantumarticle}
\pdfoutput=1
\usepackage[utf8]{inputenc}
\usepackage[english]{babel}
\usepackage[T1]{fontenc}
\usepackage{amsmath}
\usepackage{hyperref}
\usepackage{physics}
\usepackage{dsfont}
\usepackage{float}

\newcounter{pnum}
\usepackage{tikz}
\usepackage{lipsum}

\usepackage{xcolor}

\begin{document}

\title{Tomographic reconstruction of free-electron quantum states}

\author{Hao Jeng}
\email{hao.jeng@mpinat.mpg.de}
\orcid{0009-0009-2702-2991}
\affiliation{Department of Ultrafast Dynamics, Max Planck
Institute for Multidisciplinary Sciences, D-37077 Göttingen, Germany}
\affiliation{IV. Physical Institute, University of Göttingen, D-37077 Göttingen, Germany}

\author{Claus Ropers}
\email{claus.ropers@mpinat.mpg.de}
\orcid{0009-0009-2702-2991}
\affiliation{Department of Ultrafast Dynamics, Max Planck
Institute for Multidisciplinary Sciences, D-37077 Göttingen, Germany}
\affiliation{IV. Physical Institute, University of Göttingen, D-37077 Göttingen, Germany}

\maketitle

\begin{abstract}
    We give several algorithms for reconstructing quantum states of swift electrons, using maximum likelihood estimation, Bayesian inversion, and deep learning. We apply these algorithms to data previously recorded for an attosecond electron pulse-train to retrieve the density matrix and to analyse its physical properties. Based on the reconstructed quantum state, we obtain pulse-durations of about 245as and predict a degree of coherence of 36 per cent for radiations and excitations produced by these electrons.
\end{abstract}

\section{Introduction}
The quantum nature of interactions between free electrons and light form the basis of the emerging field of free-electron quantum optics. Swift electrons traversing optical near-fields can exchange an integer number of photons with the field, and be scattered into a discrete set of energy states \cite{Barwick}. When the illumination is sufficiently uniform, these energy states exist in quantum superposition \cite{feist1}. Generally, this effect is accompanied by a small number of photons being emitted spontaneously into the electromagnetic vacuum \cite{deabajo2}, creating quantum correlations between electron energy and photon number \cite{feist2, arend}. Naturally, such quantum-coherent interactions should lead ultimately to quantum entanglement between electrons and photons \cite{kfir2}, which has been observed very recently \cite{henke1, preimesberger}.

The unique features of electron-light interactions open up various new avenues of research on quantum information and technology. Laser excitation followed by dispersive free-space propagation leads to formation of attosecond electron pulse-trains \cite{feist1, priebe, morimoto, kozak}, making possible the direct imaging of attosecond dynamics at the nano-scale \cite{gaida, nabben}. Coherent cathodoluminescence \cite{polman} has been put forward as a viable method for generating novel quantum states of light \cite{dahan, digiulio}, providing a valuable source of non-linearity for quantum optics. By employing the methods of quantum metrology, entanglement between swift electrons could improve the sensitivity of electron microscopes beyond the standard quantum limit \cite{henke2, kruit}, and allow images to be taken of fragile specimens that would otherwise be easily damaged under electron-beam irradiation.

In all of the situations above, it is very important that we are able to characterise accurately the quantum states of electrons in our possession. Priebe \textit{et al.} \cite{priebe} have given a method for quantum state tomography analogous to Ramsey interferometry for atoms: the density matrix can be obtained by inducing quantum interference between the various energy states using a laser, and observing the interference patterns for each energy population with an electron spectrometer. With this method, attosecond electron pulse-trains have been synthesised and characterised with a reported pulse duration of 655as FWHM (full-width-half-maximum). Nevertheless, the reconstruction algorithm employed in those experiments is numerically involved---it is now believed that the pulse durations reported by Priebe \textit{et al.} is only a conservative estimate, and that the pulses produced in those experiments were in actuality much shorter.

In this paper we develop several algorithms for reconstructing the quantum state of swift electrons, attacking the problem from various different angles using maximum likelihood estimation, Bayesian inversion, and deep learning. These algorithms are designed to work for general quantum states, and, apart from the neural network which requires continual training, are limited in large part only by the quality of interference measurements. When applied to archival data for the attosecond electron pulse-trains recorded by Priebe \textit{et al.}, we found the electron pulses to be almost three times as short.

\begin{figure*}[t]
  \centering
  \includegraphics[width=\textwidth]{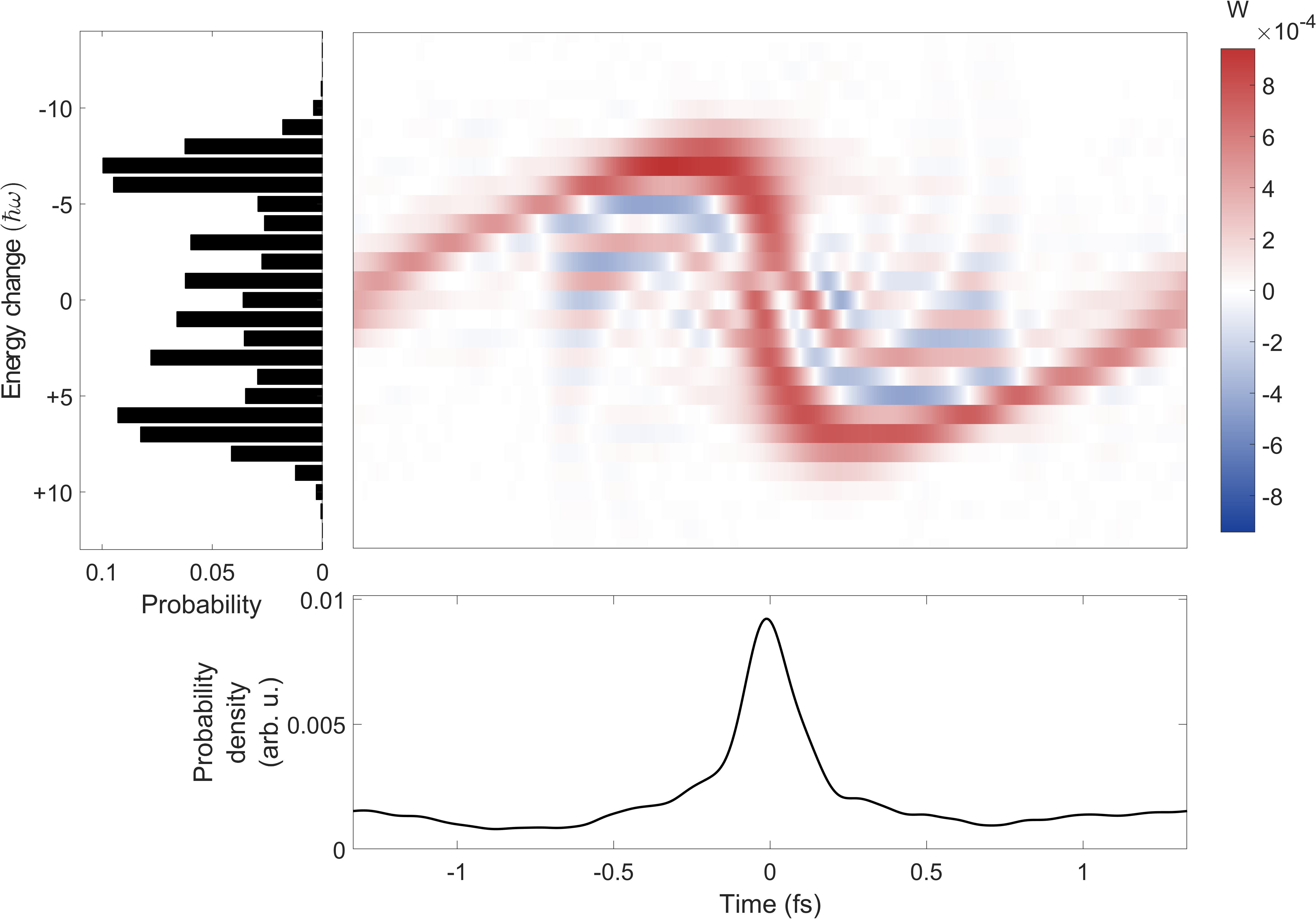}
  \caption{\textbf{Maximum likelihood reconstruction of attosecond electron pulse-trains.} Main figure shows the Wigner function calculated from the reconstructed density matrix using Eq. \ref{eq:wigner}, with the periodic temporal coordinate on the X axis and the discrete energy states on the Y axis. The temporal coordinates have been normalised to the wavelength of the laser light used (800nm), while the energies are in units of the photon energy. Both marginal distributions are shown, and attached to the relevant axes. The full-width-half-maximum of the electron pulse is 224as.}
  \label{F1}
\end{figure*}

\section{Maximum likelihood reconstruction}
To a very good approximation, quantum states of swift electrons interacting with light can be represented using a Hilbert space spanned by the discrete energy basis $\{ \ket{N} \}_{N=-\infty}^\infty$, which describes the emission and absorption of integer numbers of photons \cite{feist1}. This Hilbert space is equipped with raising and lowering operators $b^\dagger$ and $b$ which satisfy 
\begin{align}
    b^\dagger \ket{N} &= \ket{N+1}, \\
    b \ket{N} &= \ket{N-1},
\end{align}
and $[b, b^\dagger] = 0$. Interaction with an intense laser field is associated with the unitary operator 
\begin{equation} \label{eq:pinem}
    U = \exp(gb^\dagger - \bar{g}b),
\end{equation}
where the parameter $g$ is proportional to the complex amplitude of the laser and determines the strength of the interaction and its phase. The quantum state tomography procedure would then consist of an interaction $U_\phi$ with fixed intensity but varying phase $\phi$, followed by a measurement of the electron energy. For an arbitrary state $\rho$, the variation of the electron energy spectrum with the optical phase is given by the expression 
\begin{equation} \label{eq:spectrogram}
S(\phi,N) = \bra{N}U_\phi \rho U_\phi^\dagger \ket{N},
\end{equation}
which is called the ``spectrogram'' and contains all the information needed to deduce the quantum state $\rho$ \cite{shi}. In other words, the measurements described above is ``tomographically complete''.

The likelihood of obtaining a spectrogram $S$ for any given density matrix $\rho$ is described by the following expression:
\begin{equation} \label{eq:likelihood}
    L( S ~|~ \rho) = \prod_N \prod_{\phi} \bra{N} U_{\phi} \rho U_{\phi}^\dagger \ket{N}^{S(\phi,N)},
\end{equation}
and the quantum state that maximises the likelihood function could be found by fixed-point iteration with the map \cite{lvovsky}
\begin{equation}
    \rho \rightarrow \frac{R \rho R}{\Tr(R \rho R)},
\end{equation}
where the operator $R$ is defined to be
\begin{equation}
    R = \sum_{\phi,N} \frac{S(\phi,N)}{\bra{N}U_\phi \rho U_\phi^\dagger \ket{N}}  U_\phi^\dagger \ket{N}\bra{N}U_\phi.
\end{equation}
By construction, the quantum state is physical at every step of the iteration, thus offering an advantage over other numerical procedures such as semidefinite programming that require additional regularisation.

It will be useful to represent reconstructed quantum states visually by means of Wigner functions, since the features of Wigner functions tend to appear more distinctive to the eye. In contrast to previous uses of Wigner functions \cite{feist1, priebe}, however, we will use a simplified definition that reduces the phase space describing longitudinal momentum and position (or equivalently, energy and time) from $\mathds{R} \cross \mathds{R}$ to $\mathds{Z} \cross S^1$ \cite{berry, mukunda}, where $\mathds{R}$, $\mathds{Z}$, and $S^1$ denote the real numbers, integers, and the unit circle respectively. This is made possible by the quantised interaction with light which restricts energy states to a discrete subset, and imposes periodic conditions corresponding to optical cycles on the temporal densities of electrons.

With normalised position coordinates $x$ in $S^1$ and momentum $p$ in $\mathds{Z}$, the Wigner function is defined as
\begin{equation} \label{eq:wigner}
    W(x,p) = \frac{1}{2\pi} \int_{-\pi}^\pi dy \bra{x - \frac{y}{2}} \rho \ket{x + \frac{y}{2}} e^{ipy}.
\end{equation}
Position and momentum eigenstates are denoted by $\ket{x}$ and $\ket{p}$ respectively, and obey the standard relations of Fourier analysis on bounded domains \cite{stein}:
\begin{align}
    \ket{p} &= \frac{1}{\sqrt{2\pi}} \int_{-\pi}^\pi e^{ipx} \ket{x} dx,\label{eq:fourier1}\\
    \ket{x} &= \frac{1}{\sqrt{2\pi}} \sum_{p = -\infty}^\infty e^{-ipx} \ket{p}.\label{eq:fourier2}
\end{align}
Under this definition, the Wigner function is normalized as 
\begin{equation}
    \frac{1}{2\pi} \int_{-\pi}^\pi dx \sum_{p=-\infty}^\infty W(x,p) = 1,
\end{equation}
and reduces to the correct marginals 
\begin{align}
    \frac{1}{2\pi} \int_{-\pi}^\pi dx ~W(x,p) &= \bra{p} \rho \ket{p}, \\
    \sum_{p = -\infty}^\infty W(x,p) &= \bra{x} \rho \ket{x}.
\end{align}
Many quantum states in the literature appear in a much simpler form when the definition above is used. For instance, the states that maximise the coherence of cathodoluminescence \cite{yalunin} have Gaussian Wigner functions and correspond to minimum uncertainty states \cite{carruthers}.

Figure 1 shows the Wigner function of reconstructed attosecond electron pulse-trains when the maximum likelihood algorithm is applied to archival data (\cite{priebe}), with the location of the electron pulses translated in time to coincide with the midpoint of the optical cycle. In this representation, the periodic acceleration and deceleration of the electrons by the laser field is very prominent, and so is the quantum-mechanical nature of the interaction. The variation of electron energy in time is not exactly sinusoidal owing to dispersive free-space propagation, which causes an accumulation of positive probability density in the centre and the formation of attosecond pulses. The full-width-half-maximum is determined to be $8.6\%$ of the optical period, which for 800nm light corresponds to 224as.

\section{Bayesian inversion}
Whilst the maximum likelihood approach is very simple and intuitive, it is difficult to estimate meaningfully the uncertainties of the reconstructed quantities. This problem can be solved with Bayesian inversion techniques. Using Bayes' theorem, the posterior distribution of density matrices $\pi$ can be written in terms of the likelihood function and the prior $\pi_0$:
\begin{equation} \label{eq:posterior}
    \pi(\rho) \propto L(S ~|~ \rho) ~\pi_0(\rho).
\end{equation}
Since the likelihood function (Eq.~\ref{eq:likelihood}) can be evaluated straightforwardly, if we specify also a prior distribution (typically just a uniform distribution) then the posterior distribution can also be regarded as known.

In practice, however, the posterior distribution has a very large number of dimensions because electrons are scattered into many energy states, thus it is necessary to analyse the posterior distribution by means of Markov chain Monte Carlo and the ergodic theorem. The methods for dealing with this problem was first developed by Blume-Kohout \cite{blume-kohout} using the well-known random-walk Metropolis-Hastings algorithm, with further refinements made by Lukens \textit{et al.} \cite{lukens} using the pre-conditioned Crank-Nicholson proposal (pCN) which exhibits mesh-independence and provides improved scaling with the number of dimensions \cite{cotter}. The large number of dimensions remains an issue, however, as the autocorrelation of chains decay very slowly \cite{kim}, so that long computation times and substantial storage space are required to obtain statistical meaningful results.

\begin{figure*}[t]
  \centering
  \includegraphics[width=\textwidth]{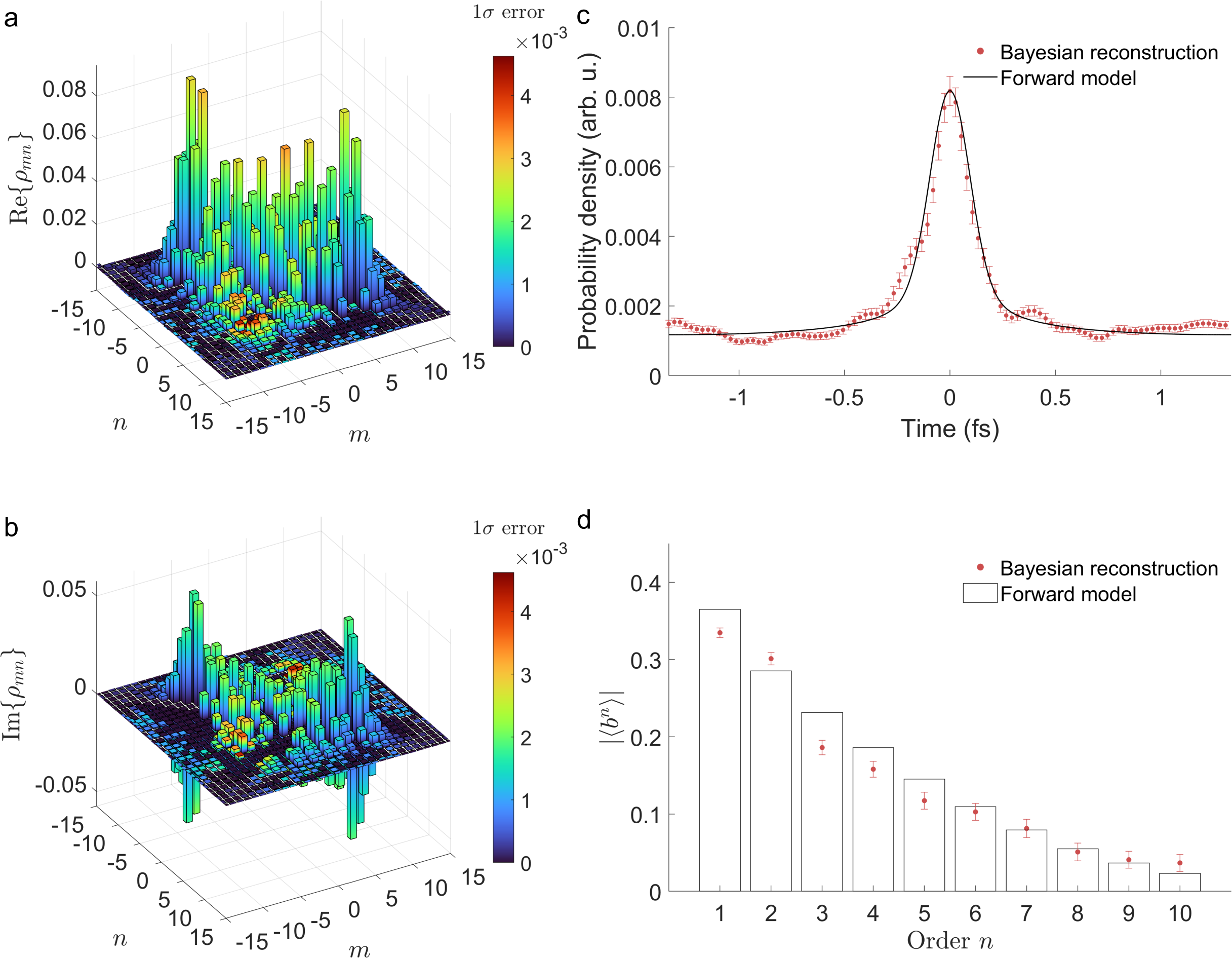}
  \caption{\textbf{Bayesian inversion.} \textbf{a} \& \textbf{b} Real and imaginary parts of the reconstructed density matrix, with the mean value of each entry represented by the bar height and uncertainties by the colour. Labels on the X and Y axes indicate changes in the electron energy in units of photon energy. \textbf{c} Temporal density, showing mean values for the Bayesian reconstruction with 1$\sigma$ uncertainty to either side, and a forward model fitted to the mean posterior density matrix (see main text for details of the model). The pulse-duration is 245(39)as. \textbf{d} The degrees of optical coherence, $|\expval{b^n}|$, expected for cathodoluminescence from these attosecond pulses.}
  \label{F2}
\end{figure*}

To carry out Bayesian inversion in a more efficient manner, we make use of a curvature-informed pCN proposal in conjunction with the Metropolis-Hastings update rule. Given sufficient measurements, the posterior distribution will typically be peaked around a certain outcome (the maximum a posteriori solution, MAP), so we will be able to draw samples from the distribution more efficiently by taking into account the curvature at this point. Since the posterior distribution can be defined algorithmically using only basic computer operations (Eq.~\ref{eq:posterior}), both of these quantities can be found using automatic differentiation \cite{wengert}: the MAP solution is obtained by gradient descent, and the Hessian matrix at the MAP point calculated using two passes of automatic differentiation for each pair of parameters. To carry out computations involving density matrices, we use a vector parameterisation \cite{chapman} that is surjective (though not injective) on the space of physical states, such that the mathematical constraints defining the set of physical density matrices is automatically accounted for. When the components of the vector are independent and normally distributed, the corresponding density matrices are uniformly distributed with respect to the Bures metric.

Concretely, the proposal function is taken to be a normal distribution \cite{pinski}:
\begin{equation}
    q(x | x_j) =  \mathcal{N}(x_{\text{MAP}} + \sqrt{1-\beta^2}(x_j - x_{\text{MAP}}), \beta^2 H^{-1}),
\end{equation}
where $x_{\text{MAP}}$ denotes the maximum a posteriori solution, $H$ is the Hessian matrix at this point, and $\beta$ is an adjustable parameter between zero and one. The state of a Markov chain is updated according to the Metropolis-Hastings rule: given a chain in the state $x_j$, a proposal $x$ is generated according to distribution above and accepted with probability 
\begin{equation}
    \alpha = \min\left\{1, \frac{\pi(x)q(x_j | x)}{\pi(x_j)q(x|x_j)}\right\}.
\end{equation}
If the proposal is accepted then we set $x_{j+1} = x$, otherwise $x_{j+1} = x_j$.

We have applied the Bayesian algorithm to reconstruct attosecond electron pulse-trains, obtaining pulse durations of 245(39)as in good agreement with the maximum likelihood approach (Fig.~\ref{F2}). The source of uncertainties is most readily recognised when representing the reconstructed density matrix directly in the energy basis (Fig.~\ref{F2}a \& b). It can be seen that the entries with the greatest uncertainties are those at the corners of the matrix corresponding to off-diagonal elements with the greatest energy differences. The reason for these uncertainties can be traced back to the interference technique for quantum state tomography: if the laser interaction is not strong enough to induce interference between a pair of energy states, then it will be difficult to deduce the quantum coherence between those states. Overall, however, the uncertainties are relatively small when compared to the most dominant entries.

We have fitted the posterior mean density matrix with a forward model that accounts for various sources of decoherence, by minimising the Frobenius norm between density matrices. At the optimum point we find a fidelity of 0.837, with a mixture of interaction strengths between $g=3.73$ and $g=4.52$ (assumed to be uniformly distributed in this range), free space propagation distance of 1.7mm, and 6.4\% normally distributed phase noise (normalised to the optical period). The results of the fitting procedure do not appear to be influenced by the initial values. Despite the indirect optimisation to attosecond pulses, the temporal profile of the forward model shows excellent agreement with that obtained using Bayesian estimation.

Free electrons that are localised in space and time generate coherent radiations \cite{kfir} and excitations \cite{gover, zhao}, with a degree of coherence captured by the quantities $\expval{b^n}$ . For example, to first order, the absolute value of $\expval{b}$ corresponds to the interference visibilities one would obtain when interfering coherent cathodoluminescence with an optimised external local oscillator. The theoretical maximum value for attosecond pulses produced by a single laser interaction and free space propagation is about 0.58 \cite{digiulio2}. We can make predictions of these quantities based on the reconstructed density matrices, and we find a degree of coherence that reaches about 63$\%$ of the theoretical maximum (Fig.~\ref{F2}d). The values obtained using the forward model agree quite well with those obtained with the Bayesian method.

\begin{figure*}[t]
  \centering
  \includegraphics[width=\textwidth]{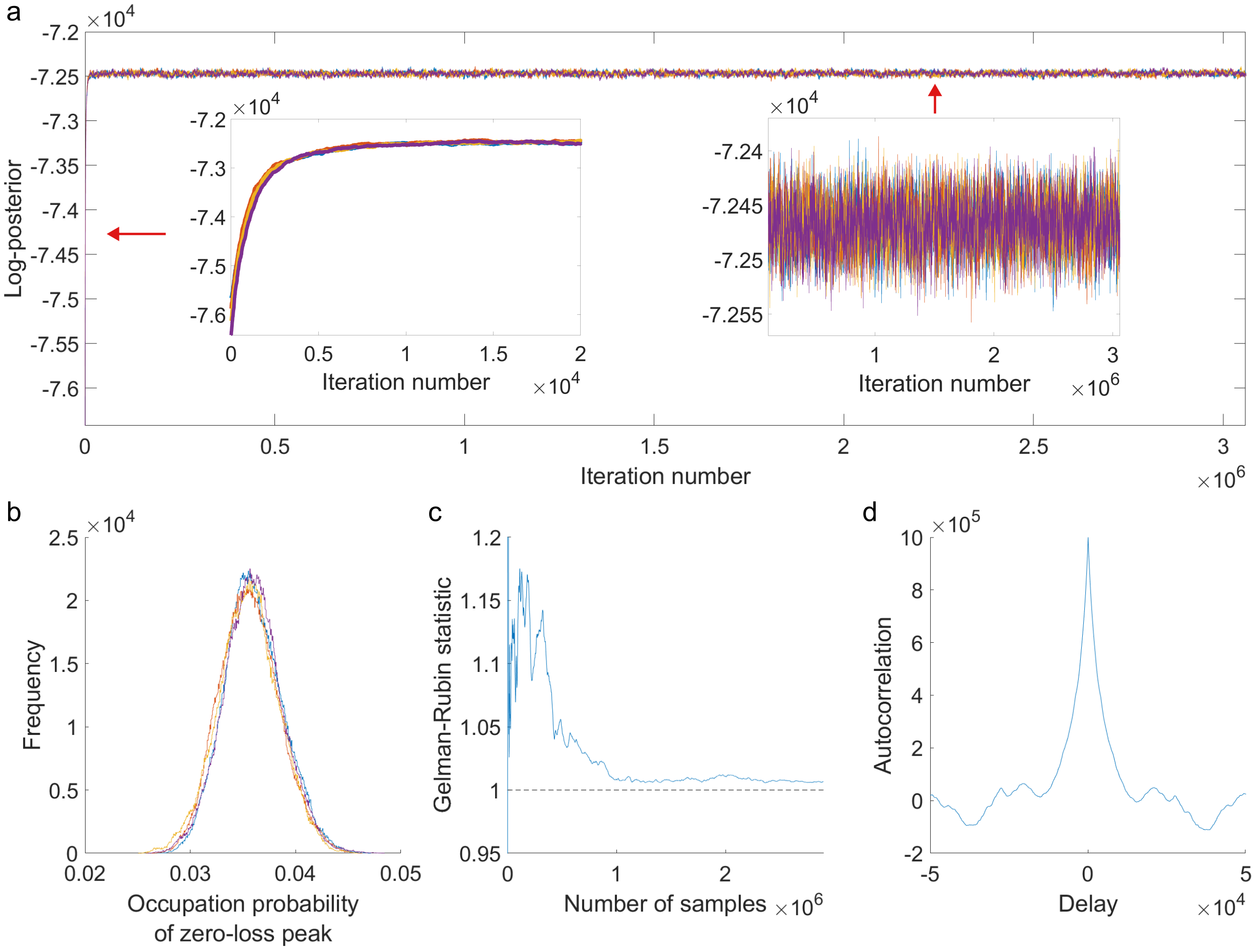}
  \caption{\textbf{Markov chain Monte Carlo.} \textbf{a} Dependence of the log-posterior on the iteration number for each of the four chains, labelled by colour, with close-up views of the first 20,000 samples and of those from 100,000 onwards. \textbf{b} Distributions of the occupation probability of the zero-loss peak for each of the four chains, labelled by colour. \textbf{c} \& \textbf{d} The Gelman-Rubin statistic and autocorrelation of chains calculated with the data for \textbf{b}.}
  \label{F3}
\end{figure*}

Figure \ref{F3} illustrates the process of sampling from the posterior distribution with Markov chain Monte Carlo. To verify convergence of the Markov chains, we generated four independent chains, with the first sample of each chain drawn randomly from the posterior distribution approximated as the multivariate normal distribution $\mathcal{N}(x_{\text{MAP}},H^{-1})$, i.e. the Gaussian approximation. Figure \ref{F3}a illustrates how these chains wander toward the MAP solution, and within 20,000 samples become centred around this point; the total number of samples for each chain was about 3 million. Generation of the four chains took place in parallel over about 2 days (AMD Ryzen 7 5800X, 32G RAM), requiring approximately 320G on an SSD to store all the density matrices that were generated.

We can check for convergence either by directly observing the distributions of each parameter of the model, with an example of the occupation probability of the zero-loss peak given in Figure \ref{F3}b, or by using the Gelman-Rubin statistic as shown in Figure \ref{F3}c; both methods indicate that relatively good convergence has been achieved. The convergence of chains for the curvature-informed pCN proposal is strongly influenced by the $\beta$ parameter, which was taken to be $0.02$ and roughly adjusted such that the acceptance probability is relatively large (0.45) while the autocorrelation is sufficiently small (Fig. \ref{F3}d).

\begin{figure*}[]
  \centering
  \includegraphics[width=0.6\textwidth]{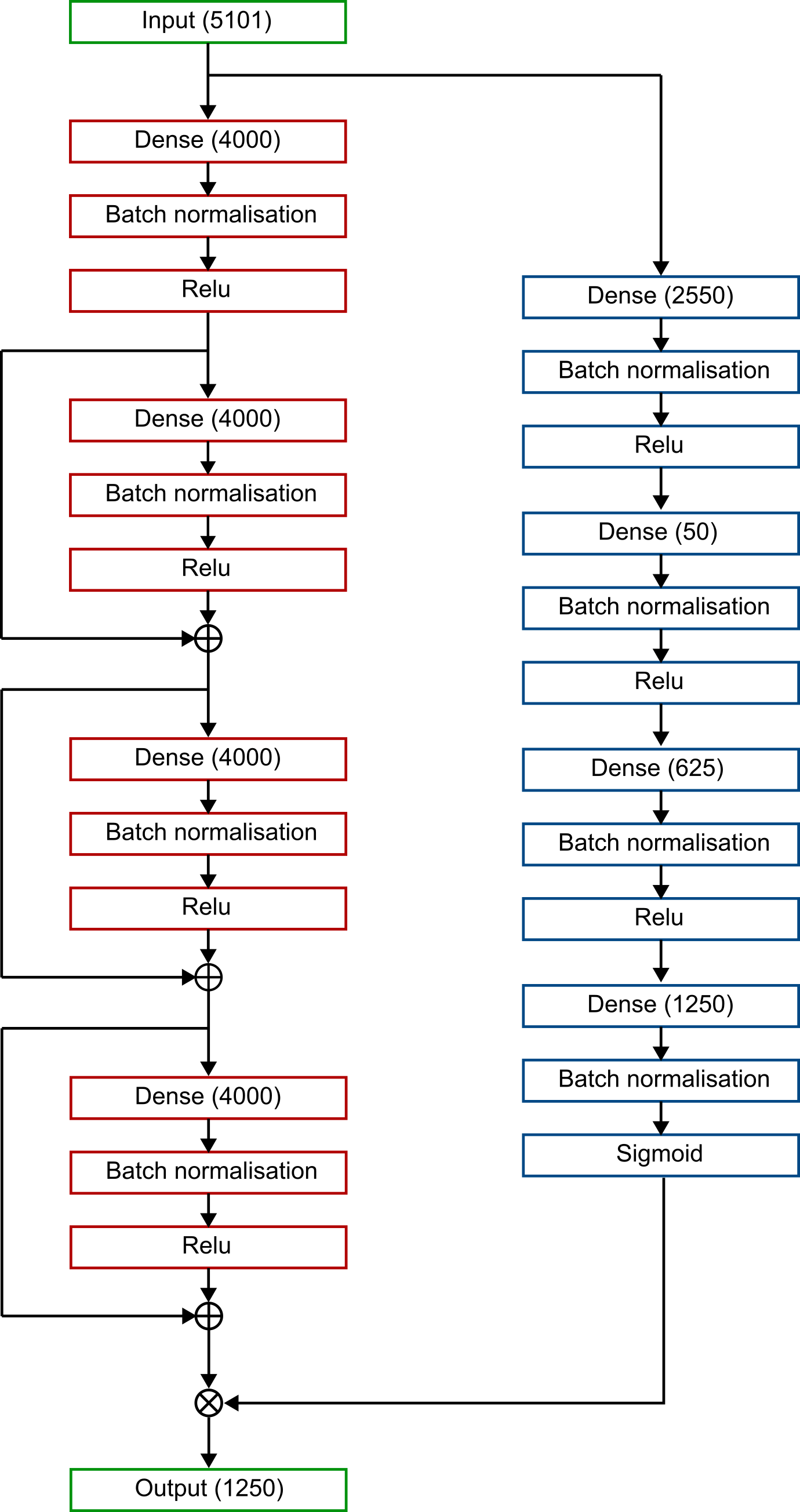}
  \caption{\textbf{Structure of neural network.} Red layers denote the primary multilayer feedforward network used for recognising quantum states, and blue layers denote an auxiliary autoencoder network used to filter noise. Detailed descriptions of each component contained in the main text.}
  \label{F4}
\end{figure*}

\section{Deep learning}
The maximum likelihood and Bayesian methods described in the previous sections are iterative algorithms and therefore can be quite slow, because these algorithms cannot be easily parallelised and thus use only a small fraction of the computing power that is available. In the remainder of this paper we develop a fast reconstruction method using deep learning, such that quantum states are reconstructed with one call to an artificial neural network. The universal approximation theorem \cite{hornik} indicates that the mapping from spectrograms to density matrices can be approximated by a neural network chosen suitably.

With some trial and error, we arrived at the structure seen in Figure \ref{F4}. It consists of a multilayer feedforward network for identifying quantum states and an autoencoder operating in parallel for filtering noise. Both networks have input layers with 5101 neurons, corresponding to spectrograms with 51 energy states and 100 phase values plus one additional parameter being the laser interaction strength, and output layers with 1250 neurons which correspond to the real and imaginary components of density matrices with 25 energy states. The multilayer feedforward network consists of four fully-connected layers with 4000 neurons each, and, to facilitate the training of this network, we found it helpful to apply batch normalisation after each layer and prior to the non-linear element (rectified linear units), as well as to make use of the techniques of residual learning by feeding forward the outputs of each layer to the outputs of the next layer. Generally, the outputs of this network is not necessarily a physical density matrix, but it is not too difficult to numerically enforce Hermiticity and positive semi-definiteness, and to normalise the resulting matrix to unit trace.

\begin{figure*}[t]
  \centering
  \includegraphics[width=\textwidth]{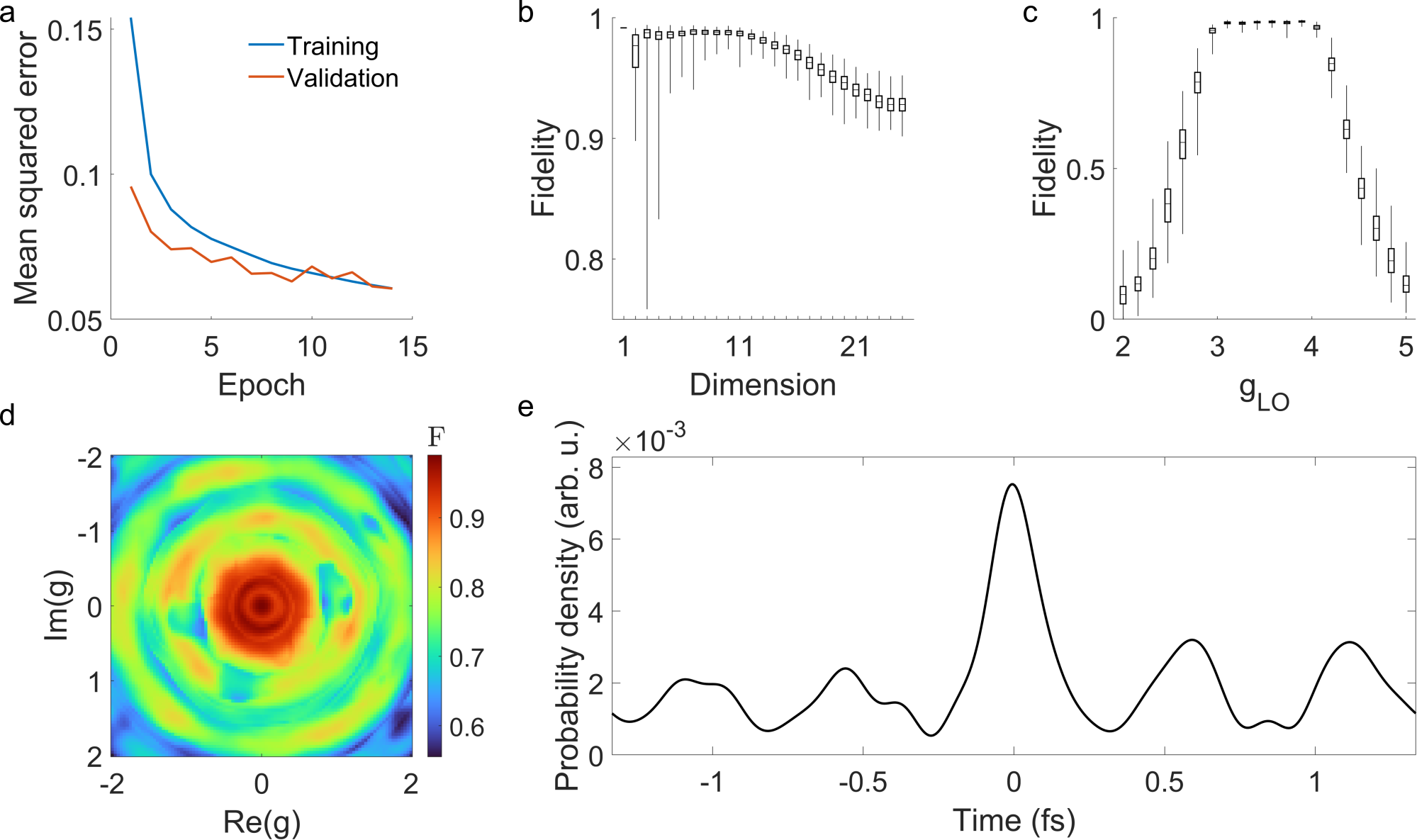}
  \caption{\textbf{Quantum state reconstruction with the neural network.} \textbf{a} Training and validation error. \textbf{b} Reconstruction fidelities for randomly generated density matrices of various dimensions (1000 samples for each dimension) visualised using a box plot. \textbf{c} Reconstruction fidelities of randomly generated density matrices with dimension of 10, and with various laser interaction strengths $g_{LO}$ for the simulated spectrograms (1000 samples for each value of the interaction strength). \textbf{d} Reconstruction fidelities, visualised as a heat map, for states of electrons obtained by a single laser excitation of a mono-energetic electron beam and without dispersive propagation (Eq. \ref{eq:pinem_state}), with various interaction strengths and phases encoded in the complex parameter $g$. \textbf{e} Temporal density of the reconstructed attosecond pulse-train. The pulse-duration is 203as FWHM.}
  \label{F5}
\end{figure*}

The autoencoder has a bottleneck of 50 neurons which corresponds to twice the maximum dimensions of the density matrix (to allow for independent control of the real and imaginary parts), and uses rectified linear units for activation in all layers except the output layer, which uses the sigmoid function. The output layer of the autoencoder is multiplied with the outputs from the multilayer feedforward network, so that the latter could be switched on and off depending on whether the sigmoid neurons are activated. The entire network, with both the multilayer feedforward network and the autoencoder, consists of approximately 87.5 million trainable parameters.

We have compiled a set of training data for the neural network, containing 20 million random density matrices up to a dimension of 25, with spectrograms simulated according to the ideal interaction Eq. \ref{eq:spectrogram}. For each sample, the laser interaction strength is drawn randomly from a uniform distribution between 3 and 4, to accommodate for variations between experiments. In addition, we found it important that the random density matrices are not generated for a Hilbert space of fixed dimension, but rather generated for a subset of randomly selected energy states and then embedded into the larger Hilbert space by padding all other energy states with zeroes. In this way, the neural network will be able to see all the lower-dimensional states as well during the training process.

The neural network was implemented using Keras with 32 bit precision and trained on a graphics processing unit (RTX 4070) using stochastic gradient descent (Adam, learning rate = 0.0001). We use mini-batches with 32 samples, mean-squared-error as the cost function, and 1\% of the training data for validation. To reduce over-fitting, the training process was terminated after 14 epochs when the training loss was very close to the validation loss (Fig. \ref{F5}a). Each epoch took approximately one hour to complete. While we have not rigorously benchmarked the execution times of each algorithm reported in this paper, a single evaluation of the neural network is on the order of 10ms, while maximum likelihood estimation typically takes about 1s.

We have tested the neural network by evaluating it for many different states, and observing that the predicted density matrices are in close agreement with the truth. Figure \ref{F5}b illustrates reconstruction fidelities for random density matrices (not taken from the training set, but generated independently), which appears to be constantly around 98\% for states of up to about 10 dimensions, at which point the fidelity begins to fall owing to the limited laser interaction strength. Figure \ref{F5}c shows reconstruction fidelities for random density matrices with a fixed dimension of 10 but with various laser interaction strengths, illustrating how the absence of training data for values of the interaction strength outside the interval between 3 and 4 causes the fidelity to fall quickly.

Figure \ref{F5}d shows reconstruction fidelities for the class of quantum states obtained by a single laser excitation of monoenergetic electrons and without dispersive propagation:
\begin{equation} \label{eq:pinem_state}
    U\ket{0} = \sum_{n = \infty}^\infty J_n(2|g|)e^{in\arg(g)} \ket{n},
\end{equation}
where $J_n$ are Bessel functions of the first kind. These states are frequently encountered in experiments, but have not been explicitly included in the training set. It is found that very high fidelities could be obtained for small interaction strengths, decaying as the interaction strengths become greater. The current state of the autoencoder appears to be cutting off energy states quite aggressively such that the states above are not reconstructed as accurately as one might expect, because the probabilities decay over some energy range and do not generally experience sharp transitions. This effect explains also the radial oscillations, as occupation probabilities oscillate around zero as a function of $g$ due to the oscillatory behaviour of Bessel functions---reconstructions are more accurate when those probabilities are closer to zero.

We have tested the neural network on the experimental data for attosecond electron pulse-trains, with the reconstructed temporal density shown in Figure \ref{F5}e. While the reconstructed quantum state is quite noisy (the energy spread being about four times that of the interaction strength $|g|$), the attosecond pulse is clearly visible, with a FWHM of 203as which more or less agrees with the other two methods. The fidelity with respect to the Bayesian posterior mean is 0.66. Owing to the large number of trainable parameters compared to the size of the training set, we can expect further refinements to these predictions by incorporating more independent samples into the training set, and with continual training of the neural network.

\section{Conclusion}
It appears quite clear to us in which situations each algorithm will be most useful: the maximum likelihood is a simple and flexible approach that can be used in most situations, while the Bayesian approach is very slow but makes maximal use of the measurement data and returns a great wealth of information about the system. Meanwhile, the neural network is the least accurate but potentially the fastest, and should find use in experiments involving real-time feedback similar to those in cavity quantum electrodynamics \cite{sayrin}.

These techniques are readily applicable to various experiments in free-electron quantum optics that may involve more complicated quantum states, such as optimised attosecond pulses with reduced background density and shorter pulse-durations \cite{deabajo, yalunin}, general superpositions of energy states in the context of quantum computing as qubits or qudits \cite{tsarev, reinhardt}, and specific types of superpositions that can be used to generate important non-classical states of light such as Schr\"{o}dinger cat states and GKP (Gottesman-Kitaev-Preskill) states \cite{dahan, digiulio}.

While we have focussed on the case of single electrons, the techniques can be extended in a straightforward manner to quantum states involving multiple electrons and photons. This is especially interesting given the recent observation of quantum entanglement between swift electrons and photons \cite{henke1, preimesberger}. It is well-known that quantum correlations between discretised energy states can be observed for electrons and photons \cite{feist2, arend}, as well as for multiple electrons \cite{haindl, meier}. Thus provided that temporal coherence between electrons and an external optical local oscillator can be established, tomographic measurements of electrons combined with those of photons (such as homodyne detection) will enable the reconstruction of these advanced quantum-mechanical systems, opening new avenues of research on quantum information and technology.

\section{Acknowledgements}
We would like to thank Katharina Priebe for making available the data for the attosecond pulses, Prof. Georg Stadler for friendly advice on Bayesian inversion, and Florian Oberender and Prof. Thorsten Hohage for their interest in our work and many enjoyable discussions. The possibility of these attosecond pulses being as short as they are was first pointed out by Thomas Rittman in independent investigations reported in an unpublished master thesis. We acknowledge financial support from the Max Planck Society and SFB 1456 (Project-ID: 432680300).

\bibliographystyle{quantum}
\bibliography{main}






\onecolumn




\end{document}